\documentclass[pss]{wiley2sp} 
\usepackage{amsmath}

\begin{document}

\title{
\begin{minipage}[t]{7.0in}
\scriptsize
\begin{quote}
\rightline{{\it Phys. Status  Solidi B}, in press}
\raggedleft {\rm arXiv:1210.0185}
\end{quote}
\end{minipage}
\medskip
Instability of the chiral phase and electronic ferroelectricity in
the extended Falicov--Kimball model}

\titlerunning{Instability of the chiral phase and Electronic Ferroelectricity}

\author{%
  D. I. Golosov\textsuperscript{\Ast,\textsf{\bfseries 1}}}

\authorrunning{Golosov}

\mail{e-mail
  \textsf{Denis.Golosov@biu.ac.il}}

\institute{%
  \textsuperscript{1}\,Department of Physics and the Resnick Institute, Bar-Ilan 
University, Ramat-Gan 52900, Israel}

\received{XXXX, revised XXXX, accepted XXXX} 
\published{XXXX} 


\abstract{%
%
%
%
\abstcol{%
We consider a spinless extended Falicov--Kimball model at half-filling, for
the case of opposite-parity bands. Within the Hartree--Fock approach, we
calculate the excitation energies in the chiral phase, which
is a possible mean-field solution in the presence of a hybridisation.
It is shown that the chiral phase is unstable.}{We then briefly review the
accumulated results on stability and degeneracies of the excitonic insulator phase. Based on these,
we conclude that the  presence of both hybridisation and 
narrow-band hopping is required
for electronic ferroelectricity.  
}}

%
%

\def\today{3 March 2013}
\maketitle   

\section{Introduction.}
In the mixed-valence regime of the extended Falicov--Kimball model 
(EFKM)\cite{Zlatic}, an
excitonic insulator state is found within a broad range of parameter 
values\cite{Batista02,Batista04,Farkasovsky08,Czycholl08,Zenker10}. When the two carrier bands involved (itinerant and nearly-localised)
have opposite parities, a spontaneous electric polarisation (electronic
ferroelectricity\cite{Portengen}) arises, provided that the off-diagonal excitonic average
value  $\langle c^\dagger_i d_i \rangle$ (induced hybridisation) is uniform and has a non-vanishing 
real part\cite{Batista02,Batista04,Portengen}. With the
experimental search for electronic ferroelectricity underway\cite{expferro}, 
it is
important to achieve a better theoretical understanding of this phenomenon.

One of the issues that needs clarification is the possibility
of the chiral phase, characterised by a uniform imaginary $\langle 
c^\dagger_i d_i \rangle = {\rm i}\Delta$ in the EFKM
with a non-zero bare hybridisation. 
This scenario was 
mentioned
earlier in the context of conventional semimetals\cite{HalperinRice}.
Further discussion of the chiral phase for the case of EFKM can be found, 
{\it e.g.},
in Ref. \cite{ZenkerBatista}. 
While not present on the available 
numerical\cite{Batista04} and 
mean-field\cite{ZenkerBatista}
phase diagrams, this phase was reported to be stable in a one-dimensional
EFKM with specific parameter values in the limit of strong interaction\cite{Sarasua}.
Below, we will address this issue within the Hartree--Fock
approximation by considering the stability of the
chiral phase with respect to collective excitations.

The issue of stability of the ferroelectric state is further complicated
by possible degeneracies of the excitonic insulator 
phase\cite{Batista02,Batista04,prl12}. In a real
system, the electrostatic dipole-dipole interaction will tend to minimise
the spontaneous polarisation $\vec{P}$, with the result that a $\vec{P}=0$ 
state will
be selected whenever such a state appears among the degenerate ground 
states. This can be easily overcome by applying an external electric field 
$\vec{E}$, yet the net result is that spontaneous polarisation vanishes
exactly (as opposed to a ferroelectric, where it may vanish only \underline{on
average} due to domain formation). This will be further discussed towards
the end of the paper. 

The Hamiltonian of the spinless Falicov--Kimball model (FKM) is given by
\begin{equation}
{\cal H}=-\frac{t}{2}\sum_{\langle i j \rangle} \left(c^\dagger_i c_j +
c^\dagger_j c_i \right) + E_d \sum_i d^\dagger_i d_i +U \sum_i c^\dagger_i
d^\dagger_i d_i c_i\,.
\label{eq:FKM}
\end{equation}
Here, the operator  $d^\dagger_i$ ($c^\dagger_i$) creates an 
electron in the localised (itinerant) band, $E_d$ is the bare energy of 
the 

\begin{minipage}[h]{16cm}
\noindent
localised electrons, and $U$ is the on-site repulsion. It will be 
assumed that the hopping amplitude $t$ and the  
%
%
 period  of the ($d$-dimensional
hypercubic) lattice are equal to unity.
We will  consider 
the half-filled case (one electron per site) at zero temperature.

While the excitonic mean field solution is present already for the pure
FKM\cite{Khomskii76}, Eq. (\ref{eq:FKM}), it is always 
unstable unless the FKM is 
\underline{extended} by a small but finite\cite{Farkasovsky08,prl12,Czycholl99} 
perturbation $\delta {\cal H}$. For the case of 
opposite-parity bands we write in the momentum space,
\begin{equation}
\delta {\cal H}=\sum_{\vec{k}}\left\{t^\prime \epsilon_{\vec{k}}
d^\dagger_{\vec{k}}d_{\vec{k}}+{\rm i}  V_2 \lambda_{\vec{k}}
\left(c^\dagger_{\vec{k}}d_{\vec{k}}-d^\dagger_{\vec{k}}c_{\vec{k}}
\right) \right\}\,,
\label{eq:pert}
\end{equation}
with  $\epsilon_{\vec{k}}=-\sum_{\alpha=1}^d \cos k_\alpha$ and
$\lambda_{\vec{k}}=-\sum_{\alpha=1}^d \sin k_\alpha$. Here, $t^\prime$
is the hopping amplitude in the narrow band, and the nearest-neighbour
hybridisation is denoted $V_2$, consistent with Ref. \cite{prl12}.  We will now
proceed with analysing the stability of the chiral state.

\section{Instability of the chiral state.}

The Hartree-Fock mean field equations for the excitonic chiral state in
the model Eq. (\ref{eq:FKM}--\ref{eq:pert}) read:
\begin{equation}
\Delta = \frac{1}{N} \sum_{\vec{k}} \Delta_{\vec{k}}^\prime\,,\,\,\,
\Delta_{\vec{k}}^\prime\equiv - {\rm i} \langle c^\dagger_{\vec{k}}d_{\vec{k}}\rangle=\frac{U \Delta+V_2 \lambda_{\vec{k}}}
{\sqrt{(\xi_{{\vec{k}}}+t^\prime \epsilon_{\vec{k}})^2+4 (U \Delta
+V_2 \lambda_{\vec{k}})^2}}
\label{eq:Delta}
\end{equation}
for the hybridisation, whereas for the narrow-band occupancy one finds:
\begin{equation}
n_d=\frac{1}{N}\sum_{\vec{k}} n^{d\,\prime}_{\vec{k}}\,,\,\,\,
n^{d\,\prime}_{\vec{k}} \equiv \langle d^\dagger_{\vec{k}}d_{\vec{k}}\rangle= 
\frac{1}{2}-\frac{\xi_{{\vec{k}}}+t^\prime \epsilon_{\vec{k}}}{2
\sqrt{(\xi_{{\vec{k}}}+t^\prime \epsilon_{\vec{k}})^2+4 (U \Delta
+V_2 \lambda_{\vec{k}})^2}
}\,,
\label{eq:nd}
\end{equation}
where  
$N$ is the number of lattice sites, 
$\xi_{\vec{k}}=E_{rd} -\epsilon_{\vec{k}}$ and 
$E_{rd}=E_d + U (1- 2 n_d)$. The notation $\Delta_{\vec{k}}$ ($n^d_{\vec{k}}$) 
is reserved for the
leading-order terms in $\Delta^\prime_{\vec{k}}$ ($n^{d\,\prime}_{\vec{k}}$),
corresponding to $V_2=t^\prime=0$, and is consistent with Ref. \cite{prl12}.

The mean-field parameters $\Delta$ and $n_d$, found from Eqs. (\ref{eq:Delta}--\ref{eq:nd}),
determine the quasiparticle dispersion in the filled and empty bands,
\begin{equation}
\varepsilon^{(1,2)}_{\vec{k}}=\frac{1}{2}
\left\{E_d+(1+t^\prime)\epsilon_{\vec{k}}+U \mp \sqrt{(\xi_{\vec{k}}+t^\prime\epsilon_{\vec{k}})^2
+(4U\Delta+V_2\lambda_{\vec{k}})^2} \right\} \,.
\label{eq:qparticle}
\end{equation}
Since $\lambda_{\vec{k}}=-\lambda_{-\vec{k}}$,
we  see that in 
the chiral phase, $\varepsilon^{(1,2)}_{\vec{k}} \neq \varepsilon^{(1,2)}_{-\vec{k}}$, 
due to the symmetry breaking\cite{ZenkerBatista}. 
We follow Ref. \cite{prl12} in writing
\begin{eqnarray}
{\cal X}_{\vec{q}}=&&\frac{1}{\sqrt{N}} \sum_{\vec{k}}\left\{
F_+(\vec{k},\vec{q})\, c^\dagger_{\vec{k}}d_{\vec{k}+\vec{q}}+
F_-(\vec{k},\vec{q})\, d^\dagger_{\vec{k}}c_{\vec{k}+\vec{q}}
+F_c(\vec{k},\vec{q})\, c^\dagger_{\vec{k}}c_{\vec{k}+\vec{q}}+
F_d (\vec{k},\vec{q})\,d^\dagger_{\vec{k}}d_{\vec{k}+\vec{q}} \right\}\,
\label{eq:wave}
\end{eqnarray}
for a generic particle-hole excitation, and perform Hartree--Fock decoupling
in the secular equation, $[{\cal X}_{\vec{q}}, {\cal H} + \delta{\cal H} ] {=}
\omega_{\vec{q}} {\cal X}_{\vec{q}}$. This yields four linear equations for
the amplitudes $F_i(\vec{k},\vec{q})$, \underline{viz.,}
\begin{eqnarray}
\!\!\!\!\!&&{\rm i}(\omega - {\xi}_{\vec{k}}- t^\prime 
\epsilon_{\vec{k}+\vec{q}}) F_+(\vec{k},\vec{q})+(U {\Delta}+
V_2\lambda_{\vec{k}+\vec{q}})F_c(\vec{k},\vec{q})
-(U {\Delta}+V_2\lambda_{\vec{k}})
F_d(\vec{k},\vec{q})=U\left[A_a(\vec{q})+A_b(\vec{q})\right]\,,
\label{eq:F+}\\
\!\!\!\!\!&&{\rm i}(\omega + {\xi}_{\vec{k}+\vec{q}}+ t^\prime 
\epsilon_{\vec{k}}) F_-(\vec{k},\vec{q})+(U {\Delta}+
V_2\lambda_{\vec{k}})F_c(\vec{k},\vec{q})
-(U {\Delta}+V_2\lambda_{\vec{k}+\vec{q}})
F_d(\vec{k},\vec{q})=U\left[A_a(\vec{q})-A_b(\vec{q})\right],
\label{eq:F-}\\
\!\!\!\!\!&&{\rm i}(U {\Delta}+V_2\lambda_{\vec{k}+\vec{q}}) F_+(\vec{k},\vec{q})+{\rm i}
(U {\Delta}+V_2\lambda_{\vec{k}}) F_-(\vec{k},\vec{q})+
(\omega + {\xi}_{\vec{k}+\vec{q}}-\xi_{\vec{k}})F_c(\vec{k},\vec{q})
=UA_c(\vec{q})\,,
\label{eq:Fc}\\
\!\!\!\!\!&&-{\rm i}(U {\Delta}+V_2\lambda_{\vec{k}}) F_+(\vec{k},\vec{q})
-{\rm i}(U {\Delta}+V_2\lambda_{\vec{k}+\vec{q}}) F_-(\vec{k},\vec{q})+
(\omega- t^\prime \epsilon_{\vec{k}+\vec{q}}+t^\prime 
\epsilon_{\vec{k}})F_d(\vec{k},\vec{q})=UA_d(\vec{q})\,,
\label{eq:Fd}
\end{eqnarray}
where the quantities $A_i(\vec{q})$ are defined self-consistently by
\begin{eqnarray}
A_a(\vec{q})&=&\frac{1}{2N} \sum_{\vec{p}}\left\{{\rm i}\left[F_+
(\vec{p},\vec{q})-F_-(\vec{p},\vec{q})\right] ({n}^{d\,\prime}_{\vec{p}}+
{n}^{d\,\prime}_{\vec{p}+\vec{q}}-1)+\left[ F_c(\vec{p},\vec{q})
-F_d(\vec{p},\vec{q}) \right]
({\Delta}^\prime_{\vec{p}+
\vec{q}}+{\Delta}^\prime_{\vec{p}})\right\}\,,
\label{eq:Aa}\\
A_b(\vec{q})&=&\frac{1}{2N} \sum_{\vec{p}}\left\{{\rm i}
\left[F_+(\vec{p},\vec{q})+F_-(
\vec{p},\vec{q})\right] ({n}^{d\,\prime}_{\vec{p}}+
{n}^{d\,\prime}_{\vec{p}+\vec{q}}-1)+\left[ F_c(\vec{p},\vec{q})
+F_d(\vec{p},\vec{q})\right]({\Delta}^\prime_{\vec{p}+
\vec{q}}-{\Delta}^\prime_{\vec{p}})\right\}\,,
\label{eq:Ab}\\
A_c(\vec{q})&=&\frac{1}{N} \sum_{\vec{p}}\left\{{\rm i}F_+(\vec{p},\vec{q}){
\Delta}^\prime_{\vec{p}}+{\rm i}F_-(\vec{p},\vec{q}) {
\Delta}^\prime_{\vec{p}+\vec{q}}+F_d(\vec{p},\vec{q})\left[{n}^{d\,\prime}_
{\vec{p}}-
{n}^{d\,\prime}_{\vec{p}+\vec{q}}\right]\right\}\,,
\label{eq:Ac}\\
A_d(\vec{q})&=&\frac{1}{N} \sum_{\vec{p}}\left\{-{\rm i}F_+(\vec{p},\vec{q}) 
{\Delta}^\prime_{\vec{p}+\vec{q}}-{\rm i}F_-(\vec{p},\vec{q}) {
\Delta}^\prime_{\vec{p}}-F_c(\vec{p},\vec{q})\left[{n}^{d\,\prime}_{\vec{p}}-
{n}^{d\,\prime}_{\vec{p}+\vec{q}}\right]\right\}\,.
\label{eq:Ad}
\end{eqnarray}

\end{minipage}

~~

\newpage


\noindent
We note the difference with Ref. \cite{prl12}, where the primary objective
was to study the case of real $\langle c^\dagger_i d_i \rangle$.

Solving Eqs. (\ref{eq:F+}--\ref{eq:Fd}) for  $F_i(\vec{k},\vec{q})$ 
and substituting into Eqs. (\ref{eq:Aa}--\ref{eq:Ad}) yields a system of four
homogeneous equations for $A_i(\vec{q})$. It is important to note that in 
the case of pure FKM, $V_2=t^\prime=0$, the first of these equations is 
trivially satisfied at $\omega=0$ (all the coefficients vanish), and 
the $A_a$-terms in the remaining equations disappear. 

The excitations spectrum, $\omega_{\vec{q}}$, is found from  the compatibility 
condition of the homogeneous system for $A_i(\vec{q})$ (zero determinant). We
already mentioned that in
the  $V_2=t^\prime=0$ case, the spectrum includes a branch which vanishes 
identically\cite{prl12},
$\omega_{\vec{q}}\equiv 0$. Physically, this is due to the local 
continuous degeneracy of the excitonic insulator in the pure 
FKM\cite{Zlatic,Subrahmanyam}. This degeneracy is lifted by a small 
perturbation, Eq. (\ref{eq:pert}), and 
the resultant low-energy branch
of the spectrum can be obtained by expansion to first order in $t^\prime$
and to second order in $V_2$ and $\omega$. We obtain the following equation
for the spectrum:
\begin{eqnarray}
D_\omega(\vec{q})\cdot \left( \frac{\omega}{U\Delta} 
\right)^2 &+& M_{11}(\vec{q}) \cdot D_0 (\vec{q})+
\nonumber \\
&+& V_2^2 F(\vec{q}) +  
\frac{\omega V_2 }{U\Delta} G(\vec{q})=0\,.
\label{eq:secular}
\end{eqnarray}
Here, the quantities $D_\omega(\vec{q})$ and $D_0 (\vec{q})$ are defined as in 
Ref. 
\cite{prl12} (where their momentum dependence is also discussed), whereas 
\begin{eqnarray}
M_{11} &= &\frac{t^\prime(d+ \epsilon_{\vec{q}}) }{U\Delta^2Nd}
\sum_{\vec{k}} \epsilon_{\vec{k}}
n^d_{\vec{k}}+\frac{V_2^2}{N}\sum_{\vec{k}}\frac{4U\lambda_{\vec{k}}^2}
{(\xi_{\vec{k}}^2+4U^2 \Delta^2)^{3/2}}- \nonumber\\
&&-\frac{V_2^2}{U\Delta^2 N}\sum_{\vec{k}}\frac{(\lambda_{\vec{k}+\vec{q}}-
\lambda_{\vec{k}})^2}{\xi_{\vec{k}+\vec{q}}-\xi_{\vec{k}}}n^d_{\vec{k}}.
\label{eq:M11}
\end{eqnarray}
Whilst the first term in $M_{11}$ is the same as in the case of a 
real  $\langle c^\dagger_i d_i\rangle$, considered in   Ref. \cite{prl12}, the 
terms which contain $V_2$ differ, reflecting the difference in the
hybridised bandstructure. Furthermore, two new terms arise in Eq. (\ref{eq:secular}), \underline{viz.,} the third term, containing an off-diagonal determinant
$F$ 
~[with $F(\vec{q})=F(-\vec{q})$], 
and the last term, which is proportional to $\omega V_2$ and odd in
momentum, $G(\vec{q})=-G(-\vec{q})$. The latter term is due to the imaginary
$\langle c^\dagger_i d_i\rangle$ breaking the inversion symmetry of the 
quasiparticle bands, Eq. (\ref{eq:qparticle}). This in turn leads to breaking
the symmetry of the collective excitation spectrum, $\omega_{\vec{q}}\neq
\omega_{-\vec{q}}$, as follows from  Eq. (\ref{eq:secular}). While suppressing 
the lengthy explicit expressions for $F(\vec{q})$ and $G(\vec{q})$, we note 
that
both quantities contain integrals of the form
\[
\frac{1}{2N}\sum_{\vec{k}}\frac{\lambda_{\vec{k}+\vec{q}}-
\lambda_{\vec{k}}}{\xi_{\vec{k}+\vec{q}}-\xi_{\vec{k}}}\left[ f(\xi_{\vec{k}+\vec{q}})-f(\xi_{\vec{k}})\right]\,,
\]
and therefore vanish at $\vec{q} \rightarrow 0$, as does the last term in Eq. (\ref{eq:M11}).

For our immediate purposes, it is sufficient to notice that at $\vec{q}=0$ one obtains
\begin{equation}
\left(\frac{\omega_{\vec{q}}}{U\Delta}\right)^2=-\frac{D_0(\vec{q})}{D_\omega(\vec{q})} \cdot
\frac{4UV_2^2}{N}\sum_{\vec{k}} \frac{\lambda_{\vec{k}}^2}{(\xi_{\vec{k}}^2+4U^2V_2^2)^{3/2}} <0\,,
\label{eq:instab}
\end{equation}
where we used the fact\cite{prl12} that both $D_0$ and $D_\omega$ are negative at $\vec{q}=0$. 
Equation (\ref{eq:instab}), which holds whether $t^\prime$ vanishes or not, guarantees that the chiral state
is unstable at small $V_2 \neq 0$ 
(this implies that the mean-field solution,
corresponding to the chiral state, 
represents a saddle point rather than a local energy minimum). 

The origin of this instability is  different from that of the instability which arises\cite{prl12} in the excitonic insulator state of the EFKM
 in the leading order in $t^\prime<0$.
The latter instability is due to a sign change in $D_0(\vec{q})$ as $\vec{q}$ varies from $\vec{q}=0$ to $\vec{q}=
\{\pi,\pi(,\pi)\}$; taking into account the next-order (in $t^\prime$) terms leads to a substitution $D_0 \rightarrow D_0+D_1$.
When the value of $|t^\prime|$ is not too small, $-t^\prime> |t^\prime_{\rm cr}|$, the quantity $D_0(\vec{q})+D_1(\vec{q})$
 remains negative
for all $\vec{q}$, resulting in a stabilisation of the excitonic insulator at
$V_2=0$. We note that in this $V_2=0$, $t^\prime < 0$  case
the calculations of Ref. \cite{prl12} remain valid whether $\langle c^\dagger_i d_i\rangle$ is real or imaginary, owing
to exact symmetry of the EFKM mean-field Hamiltonian with respect to the 
phase of $\langle c^\dagger_i d_i\rangle$ ~\cite{Batista02,Batista04}.

In the $V_2 \neq 0$ case, on the other hand, the instability of the chiral state, Eq. (\ref{eq:instab}), arises due to the sign of the $V_2$-containing terms in the system
of equations for $A_i$ (i. e.,  to $M_{11}$, rather than $D_0$, being of the ``wrong'' sign at small $\vec{q}$). Hence even
for  $-t^\prime> |t^\prime_{\rm cr}|$, when the spectrum of the excitonic insulator state with imaginary  $\langle c^\dagger_i d_i\rangle$ is stable for $V_2=0$, including an arbitrary small $V_2$ leads to an instability [since the first term
on the r. h. s. of Eq. (\ref{eq:M11}) vanishes at $\vec{q}=0$]. This situation is 
therefore expected to persist also when the higher-order  (in $V_2$) terms are taken into account. We 
conclude that within the Hartree--Fock approach, the chiral state should remain
unstable at least as long as the effects of $V_2$ can
be treated perturbatively. 

\section{Conditions for electronic ferroelectricity.}
We begin with the case of $t^\prime<0$ and $V_2=0$. Away from the symmetric case $E_d=0$, the excitonic insulator state
is stabilised in the mixed-valence regime\cite{Batista02,Batista04,Farkasovsky08,Czycholl08}, 
provided\cite{Farkasovsky08,prl12} that $|t^\prime|$ is not too small, $-t^\prime>|t^\prime_{\rm cr}(E_d)|$. A Goldstone
mode is present at $\vec{q}=0$, corresponding to the $U(1)$ 
degeneracy of the Hamiltonian\cite{Batista02,Batista04}. The associated 
fluctuations do not destroy the long-range order in three dimensions, 
yet
the ordered state breaks the symmetry of the model, Eqs. 
(\ref{eq:FKM}--\ref{eq:pert}) with respect to the phase of the (uniform)
$\langle c^\dagger_i d_i\rangle$, and is therefore degenerate.

We will now take into account the electrostatic effects, 
including both the interaction with an external electric field $\vec{E}$ and 
the dipole-dipole
interaction. At zero temperature, the net energy takes form
\begin{equation}
{\cal H} +  
\delta {\cal H} -\int_{\rm (sample)} \left(\vec{E}\cdot \vec{P} +
\frac{1}{2}\vec{E}_{\rm int} \cdot 
\vec{P}\right) dV\,\,
\label{eq:thermod}
\end{equation}
Omitting other possible contributions (discussed in the standard 
literature), in the case of a uniform $\langle c^\dagger_i d_i\rangle$
we write for 
the polarisation density\cite{Batista02,Portengen},
\begin{equation}
\vec{P}= 2 \vec{\mu} {\rm Re}  \left( \frac{1}{N}\sum_{\vec{k}}
 \langle c^\dagger_{\vec{k}}d_{\vec{k}}\rangle \right)
\end{equation}
[when $\langle c^\dagger_i d_i\rangle$ is non-uniform, straightforward 
generalisation yields a non-uniform $\vec{P}(\vec{r})$]. 
The vector $\vec{\mu}$ is the interband 
matrix element of the dipole moment operator, and its direction 
is determined by the structure of the relevant orbitals. In Eq. 
(\ref{eq:thermod}), the internal electric field $\vec{E}_{\rm int}$ is the 
dipole field created by the polarised sample itself, hence the factor 
$1/2$ to avoid double 
counting; inside a uniformly polarised ball, $\vec{E}_{\rm int}
=-4 \pi\vec{P}/3$. This field differs from zero also outside the sample, and 
whether $\vec{P}$ is uniform or not, the
contribution of the last term in the integrand in Eq. (\ref{eq:thermod}) to the total energy is
\begin{equation}
-\frac{1}{2}\int_{\rm (sample)} \vec{E}_{\rm int} \cdot 
\vec{P} dV = \frac{1}{8\pi} \int_{\rm (entire \,\,\,  space)} {E}_{\rm int}^2 dV
\end{equation}
-- a positive quantity, which for the case of a uniformly polarised sample
is proportional to $P^2$. 

At $\vec{E}=0$, ferroelectrics\cite{Burfoot} (like
ferromagnets at zero magnetic field\cite{Aharoni}) tend to subdivide into 
domains
in order to reduce the value of this term (which nevertheless remains
positive) at the cost of slightly increasing the contribution of the 
microscopic part  [in our case, ${\cal H} +  
\delta {\cal H}$]; the latter increase is due to domain walls formation.
However, presently we encounter a different type of behaviour. As we shall see
momentarily, it will suffice to consider the case of uniform polarisation
$\vec{P}$.

In addition, we may assume that the electrostatic terms in 
Eq. (\ref{eq:thermod}) 
represent a small correction,
so that generally one would first minimise the value of the microscopic 
energy,  Eqs. (\ref{eq:FKM}--\ref{eq:pert}), and then calculate the 
value of these
terms for the appropriate ground state. However, in the present case,
owing to the degeneracy of the
microscopic Hamiltonian, the true ground state is chosen from
the manifold of (microscopically) degenerate ground states  (which 
are characterised by different values of $P$)
by minimising the electrostatic contribution.

When $\vec{E}=0$, we see that (as mentioned in the Introduction) the
dipole-dipole interaction
will stabilise the state with a uniform imaginary $\langle c^\dagger_i d_i\rangle$, corresponding to zero polarisation.  
The electrostatic contribution to
the energy then vanishes altogether, while the microscopic energy retains
its minimal value. 
Applying a small electric field $\vec{E}$ would suffice to make $\langle c^\dagger_i d_i\rangle$ real and saturate 
$\vec{P}$. 
For a spherical
sample, polarisation saturates at
$|\vec{E}\cdot \vec{\mu}/\mu| > E_{\rm cr}$, where [cf. Eq. 
(\ref{eq:thermod})] $E_{\rm cr}= 8 \pi |\mu| 
| \langle c^\dagger_i d_i\rangle |/3$. 
At $|\vec{E}\cdot \vec{\mu}/\mu|<E_{\rm cr}$,  polarisation would 
increase linearly with the
field, while $\vec{E}_{\rm int}=-(\vec{E} \cdot \vec{\mu})\vec{\mu}/\mu^2$. The system will
therefore possess a \underline{divergent dielectric constant}, 
but would not show 
ferroelectricity. 

In two dimensions the EFKM, Eqs. (\ref{eq:FKM}--\ref{eq:pert}) with $V_2=0$, does not support the excitonic long-range order
owing to the Mermin--Wagner theorem\cite{Batista04}. 
At $\vec{E}=0$, the dipole-dipole interaction will restore the
 long-range order
(again with imaginary   $\langle c^\dagger_i d_i\rangle$ , hence $\vec{P}=0$) for $-t^\prime >| t^\prime_{\rm cr}|$,  and below a certain low critical temperature $T_{\rm dd}$.

In the $t^\prime=0$, $V_2 \neq 0$ case, the excitonic insulator state can be stabilised in three dimensions (in agreement with
Ref. \cite{Portengen}), provided\cite{prl12} that $|V_2|> V_{2,{\rm cr}}(E_d)$. Moreover, a uniform imaginary  $\langle c^\dagger_i d_i\rangle$ 
corresponds to the (unstable) chiral state and therefore would not be selected. However at least within the mean field
description there is a Goldstone mode surviving  at $\vec{q}=\{\pi,\pi(,\pi)\}$, with an associated ground-state degeneracy
as discussed in Ref.\cite{prl12}. The effect of the dipole-dipole interaction at $\vec{E}=0$ will be that  $\langle c^\dagger_i d_i\rangle$  will order in a checker-board fashion, and will be imaginary: $\langle c^\dagger_i d_i\rangle=\pm {\rm i} \Delta$.
Thus, the behaviour of $\vec{P}(\vec{E})$ will be much like in the $t^\prime<0$, $V_2=0$  case 
(and $ E_{\rm cr}$ will be expressed in the same way). 
This is at variance with Ref. \cite{Portengen}, where the effects of electrostatic dipole-dipole
interaction were not discussed.

Finally, in the $V_2 \neq 0$, $t^\prime<0$ case the excitonic insulator state with real  $\langle c^\dagger_i d_i\rangle$ can be
stabilised, provided that at least one of $|V_2|$ and $t^\prime$  is 
sufficiently large\cite{prl12}.
There are no Goldstone
modes, and as demonstrated above the chiral state (with a uniform imaginary  $\langle c^\dagger_i d_i\rangle$) remains
unstable at $t^\prime \neq 0$. We conclude that in this case even with the electrostatic dipole-dipole interaction taken into
account, the EFKM shows a proper electronic ferroelectricity, characterised by a non-vanishing spontaneous polarisation,
$\vec{P}\neq 0$ at $\vec{E}=0$ . This is in agreement with earlier results\cite{Batista04}, obtained for a
two-dimensional model.

\begin{figure}[t]%
\includegraphics*[width=\linewidth,height=\linewidth]{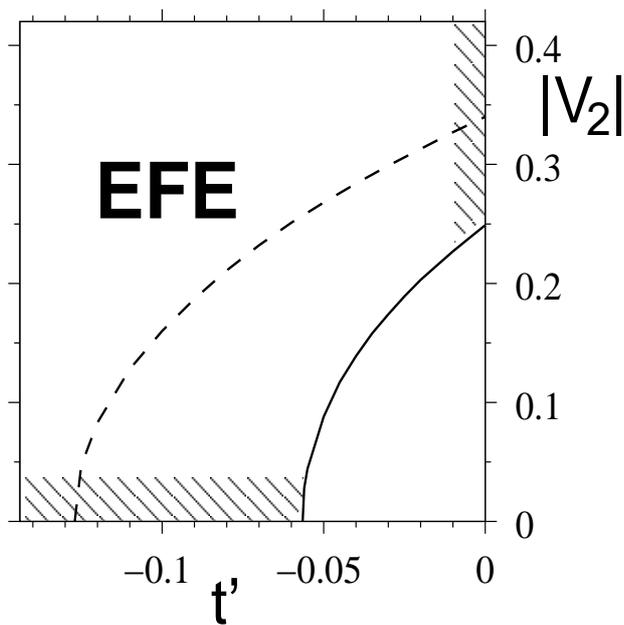}
\caption{%
Electronic ferroelectricity (EFE) in a 
two-dimensional EFKM with $U=2$ at zero temperature. 
For $E_d=0.4$ ($E_d=0.25$), EFE is 
stable above the solid (dashed) line; this line is found based on
Ref. \cite{prl12} as the phase boundary of a uniform excitonic insulator. 
When  either $|V_2|$ or $|t^\prime|$ becomes too small (shown schematically
by hatching; the actual width of these regions depends on the parameters of 
the system), 
the EFE is destabilised by the electrostatic dipole-dipole 
interaction, as discussed in the text.  
}
\label{fig:diagram}
\end{figure}

When either  $t^\prime$ or $|V_2|$ approaches zero, the dipole-dipole
interaction would eventually destabilise the ferroelectric phase
(polarisation at $\vec{E}=0$ vanishes, followed by an increase of $E_{\rm cr}$ 
towards the $t^\prime=0$ or $V_2=0$ value).
This overall behaviour is reflected in Fig. \ref{fig:diagram}, showing
the stability region for electronic ferroelectricity in a two-dimensional
EFKM with specific parameter values. For the case of three dimensions, 
similar results are expected.

\section{Summary.}

We verified that the chiral excitonic insulator state of the EFKM is unstable 
within the 
Hartree--Fock mean-field approach. Eliminating this possible candidate for
the ground state allows to answer a general question about the possibility
of electronic ferroelectricity. We find that the latter requires the presence
of \underline{both} hybridisation and narrow-band hopping. 
When only one of
these is present,  
the dielectric constant diverges,  yet the electrostatic
dipole interactions preclude the formation of a spontaneous polarisation. 
This is due to the degeneracies of the excitonic insulator
state.~

\begin{acknowledgement}
The writer takes pleasure in thanking R. Berkovits and  
K. A. Kikoin
for discussions.
This work was supported by the Israeli Absorption Ministry. 
\end{acknowledgement}



%
%

\end{document}